\documentclass[aps,prd,print,superscriptaddress,showkeywords,showpacs]{revtex4-1}
\usepackage{mathtext,bbm,amsmath,amsfonts,amssymb,indentfirst,syntonly,graphicx}
\usepackage[english]{babel}
\usepackage{amsmath,amssymb,amsthm}
\usepackage{latexsym,graphicx,bbm}

\newcommand{\beq}{\begin{eqnarray}}
\newcommand{\eeq}{\end{eqnarray}}

\newcommand{\ba}{\left( \begin{array}}
\newcommand{\ea}{\end{array} \right)}

\begin{document}

\title{Magnetized strange quark matter in a mass-density-dependent model}
\date{\today}
\author{J. X. Hou}

\affiliation{Theor. Physics Center for Science Facilities, Institute of High Energy Physics, Beijing 100049, China}

\author{G. X. Peng}
\affiliation{Theor. Physics Center for Science Facilities, Institute of High Energy Physics, Beijing 100049, China}

\affiliation{School of Physics, University of Chinese Academy of Sciences, Beijing 100049, China}

\author{C. J. Xia}
\affiliation{School of Physics, University of Chinese Academy of Sciences, Beijing 100049, China}
\author{J. F. Xu}
\email[]{xjfil@126.com}
\affiliation{School of Physics, University of Chinese Academy of Sciences, Beijing 100049, China}
\begin{abstract}
We investigate the properties of strange quark matter in a strong magnetic field with quark confinement by the density dependence of quark mass considering the total baryon number conservation, charge neutrality and chemical equilibrium. The strength of the magnetic field considered in this article is $10^{16} \sim 10^{20}$ G. It is found that an additional term should appear in the pressure expression to maintain thermodynamic consistency. At fixed density, the energy density of magnetized strange quark matter varies with the magnetic field strength. The exists a minimum with increasing the field strength, depending on the density. It is about $6\times10^{19}$ Gauss at two times the normal nuclear saturation density.
\end{abstract}

\pacs{24.85.+p, 12.38.MH, 25.75.-q, 21.65.+f, 26.60.+c}
\maketitle
\section{Introduction}

Strange quark matter (SQM) is a form of matter which contains a large quantity of deconfined quarks in $\beta$-equilibrium, with electric and color charge neutrality. The conjecture that SQM could be the true ground state of strong interactions was proposed by Bodmer-Wittern-Terazawa \cite{Bodmer1971,Witten1984,Terazawa1979}. Farhi and Jaffe \cite{Farhi1984} studied SQM in the framework of the MIT bag model \cite{Jaffe1974} for various values of the strange quark mass and the bag constant. SQM could be succeeded in the inner core of neutron stars where strange quarks would be produced through the weak processes with a dynamical chemical equilibrium among the constituents. It is possible that after a supernova explosion its core forms directly a strange quark star (SQS) \cite{Hatsuda1987,Sato1987} self-bound by strong interactions while in an general scenario neutron stars are bound by gravitational force. An astrophysical object could also form a hybrid star that has a quark core and a crust of hadronic matter.

The stability of SQM is strongly affected in a strong magnetic field \cite{Chakrabarty1996}. In a real astrophysical scenario, the strong magnetic field plays an important role. The typical strength on the surface of pulsars could be of the order $\sim 10^{12}$ G. Magnetars and neutron stars could be associated with sources with intense magnetic fields around $\sim 10^{13}-10^{15}$ G or even higher \cite{Duncan1992,Kouveliotou1988}. The origin of such ultrastrong magnetic fields could be explain in two ways. One is that the magnetohydrodynamic dynamo mechanism with large magnetic fields generated by rotating plasma of a protoneutron star \cite{Kouveliotou1988}. The other is that during the star collapse with magnetic flux conservation, the relatively small magnetic fields were amplified \cite{Tatsumi2006}. In a recent research, it was found that noncentral high-energy heavy-ion collisions could generate intense magnetic fields as high as about $10^{19}$ G, corresponding to $eB_m \sim 6m_{\pi}^2$, where $e$ is the fundamental electric charge and $m_{\pi}$ is the pion mass.

Since the lattice approach still has difficulty in consistent treatment of the finite chemical potential and the application of perturbative quantum chromodynamics (QCD) to the strong-coupling domain is unbelievable, we have to use phenomenological models in most cases. In Ref. \cite{Chakrabarty1996}, Chakrabarty studied quark matter in a strong magnetic field with conventional MIT bag model, and found that SQM becomes more stable if the order of the strength of the magnetic field is greater than some critical value. In Refs. \cite{Rojas2005,Rojas2007}, the authors confirmed that there is an anisotropy of pressures due to the strong magnetic field \cite{Chaichian2000,Rojas2003,Mosquera2005} and that the MIT bag model can be used to study magnetized SQM (MSQM)satisfactorily. With the Nambu-Jona-Lasinio (NJL) model the properties of mSQM were also discussed by many researchers \cite{Ebert2003,Frolov2010,Fayazbakhsh2011,Menezes2009,Avancini2011}. The linear sigma model coupled to quarks and to the Polyakov loop was used to investigate the influence of strong magnetic field background on confining and chiral properties of QCD \cite{Mizher2010}, where the impact of the vacuum correction from quarks on the QCD phase structure was discussed.

In nuclear physics it is well known that particle masses vary with environment. Such masses are called effective masses \cite{Walecka1995,Henley1990,Brown1991,Cohen1992}. Effective masses and effective bag constants for quark matter had been broadly studied \cite{Schertler1997,Buballa1999}.In quasiparticle model, the particle mass is derived at the zero-momentum limit of the dispersion relations from an effective quark propagator by resuming one-loop self-energy diagrams in the hard-dense-loop approximation. This reveals the dependence of particle masses on chemical potentials. In a  recent research \cite{Wen2012}, the authors extended the quasiparticle model to studying MSQM. They find a density- and magnetic-field-dependent bag function, which has a maximum at $2-3$ times the saturation density when the QCD scale parameter is larger than $123$ MeV.

In the present paper, we apply another quark model with confinement by the density dependence of quark masses. In this model the masses of quarks depend upon the baryon number density. This idea was initially introduced by Fowler, Raha, and Weiner and was used to study light quark matter \cite{Fowler1981}. Later Chakrabarty and coworkers applied the model to the case of SQM \cite{Chakrabarty1989,Chakrabarty1991,Chakrabarty1993}. The main advantage of CDDM model is the inclusion of quark confinement without using the bag constant. Instead, it is achieved by the density dependence of  the quark masses derived from in-medium chiral condensate \cite{Peng1999,Peng2000,Wen2005}. The two most  important concentrations in this model are the quark mass scaling \cite{Peng1999,Wen2005} and the thermodynamic treatment \cite{Peng2000,Wen2005}. In the beginning, the interaction part of the quark masses was assumed to be inversely proportional to the density \cite{Fowler1981,Chakrabarty1989}. Researchers also suggested other mass scalings \cite{Wang2000,Peng1997}. In Ref. \cite{Peng1999} and \cite{Wen2005}, a cubic root scaling was derived based on the in-medium chiral condensates and linear confinement at zero and finite temperature respectively. This scaling has been used to investigate many aspects of SQM \cite{Zheng2004,Lugones2003,Wen2007,Shen2007}. In the present article, we use the CDDM model with cubic root mass scaling to study the properties of SQM in a strong magnetic field.

The article is organized as follows. In Sec. \ref{sec:TT}, we derive the thermodynamic formulas when quark masses are density dependent. In Sec. \ref{sec:posqm}, we analyze the properties of mSQM and present our numerical results. The effects of the magnetic field on the system are also discussed. In Sec. \ref{Mr}, the mass-radius relation of magnetized quark stars is investigated. A short summary is presented in Sec. \ref{conclusion}.

\section{Thermodynamics Treatment \label{sec:TT}}

In the CDDM model, quark confinement is achieved by the density dependence of quark masses: with decreasing density, the mass of a quark becomes infinitely large so that the vacuum is unable to support it. Therefore, the proper form of the density dependence is very important. Originally, the quark masses are parameterized as

\begin{equation}\label{oldscaling}
  m_{q}=m_{q0}+\frac{B}{3n},
\end{equation}
where $q=u, d, s$ quarks, $m_{s0}$ is the current mass of flavor $q$, $B$ is the famous MIT bag constant, $n$ is the baryon number density.

Based on the in-medium chiral condensates, a cubic root scaling was derived at zero temperature \cite{Peng1999}, and it has been recently extended to finite temperature \cite{Wen2005}. We adopt this mass scaling in the present article. At zero temperature, the mass scaling is

\begin{equation}\label{scaling}
  m_{q}=m_{q0}+\frac{D}{n^{\frac{1}{3}}},
\end{equation}
where $m_{q0}$ is the quark current mass, $D$ is the confinement parameter, $n$ is the total baryon number density, the exponent of density was derived based on the in-medium chiral condensates and linear confinement at zero temperature \cite{Peng1999}. In the present model, the parameters are the electron mass $m_e$, the quark current masses $m_{u0},m_{d0},m{s0}$, and the confinement parameter $D$. The electron does not participate in strong interaction, its mass is a constant, i.e. $m_{e}=0.511$ MeV. Since the light-quark current masses are too small compared to the interaction part, we take $m_{u0}=m_{d0}=0$. The strange quark current mass is $95\pm25$ MeV \cite{Yao2006}. The parameter $D$ has been discussed in Ref. \cite{Peng2000,Wen2005,Peng2008}. We treat $D$ as a free parameter here. The value of $D^{1/2}$ should be in the range $(156,270)$ MeV \cite{Peng2008,Wen2005}.

The starting point of our manipulation for SQM in a strong magnetic field is the energy density. Let's start with the energy density expression os a free particle system
\begin{equation}
  E=\sum_iE_{i}=\sum_i\frac{2 g_i}{(2\pi)^3}\int\int\int \sqrt{p^2+m_i^2}\mbox{d}^3\vec{p},
\end{equation}
where the sum is over $u,d,s$ quarks and the electron, $g_i$ is the degeneracy factor ($g_i=3$ (color) for quarks and $g_i=1$ for electrons). The degeneracy due to spin has been denoted by factor $2$ in the numerator.

We assume the  magnetic field to be directed along the $z$ axis with constant field strength $B_m$. The single particle energy spectrum is given by \cite{Laudau1965}
\begin{equation}
  \epsilon_{n,s}^{(i)}=\sqrt{p_{z}^{2}+m_{i}^{2}+e_{i}B_\mathrm{m}(2n+s+1)},
\end{equation}
where $n=0,1,2,...,$ are the principal quantum numbers for allowed Laudau levels, $e_{i}$ is the absolute value of the electronic change (e.g., $e_{u}=2/3$, $e_{d}=e_{s}=1/3$, and $e_{e}=1$), $s=\pm1$ refers to spin-up or spin-down states, and $p_{z}$ is the component of particle momentum along the direction of the external magnetic field.
We now should replace the integration over $p_{x}-p_{y}$ plane in the momentum space by the rule \cite{Chakrabarty1996}
\begin{equation}\label{rule}
  \int\int dp_{x}dp_{y}\longrightarrow2\pi e_{i}B_\mathrm{m}\sum_{s=\pm1}\sum_{n},
\end{equation}
Notice that the ($0,-1$) state is a nondegenerate state while ($n_0,1$) state and ($n_0+1,-1$) degenerate, we can simplify the notation by setting $2\nu=2n+s+1$, where $\nu=0,1,2,...$. Then the single particle energy spectrum becomes
\begin{equation}
  \epsilon_{n,s}^{(i)}=\sqrt{p_{z}^{2}+m_{i}^{2}+2\nu e_{i}B_\mathrm {m}}£¬
\end{equation}
and the substituting rule in Eq. (\ref{rule}) was switched to
\begin{equation}\label{nu}
  2\pi e_{i}B_\mathrm{m}\sum_{s=\pm1}\sum_{n}\longrightarrow2\pi e_{i}B_\mathrm{m}\sum_{\nu}(2-\delta_{0\nu}).
\end{equation}
Accordingly, the energy density of mSQM is  given by
\begin{equation}\label{EIF}
  E=\sum_i E_{i}=\sum_i \frac{g_{i}e_{i}B_\mathrm{m}}{4\pi^2}\sum_{\nu=0}(2-\delta_{\nu 0})\int \sqrt{p_{z}^2+m_i^2+2\nu e_{i}B_\mathrm{m}}\mbox{d}p_{z},
\end{equation}
Explicitly carrying out the integration
\begin{eqnarray}\label{EIZ}
  E_{i}&=&\frac{g_{i}e_{i}B_\mathrm{m}}{2\pi^2}\sum_{\nu=0}^{\nu_{\textrm{max}}}(2-\delta_{\nu 0})\int_{0}^{p_{i,\nu}} \sqrt{p_{z}^2+M_{i,\nu}^2}\mbox{d}p_{z}\\\label{EIY}
  &=&\frac{g_{i}e_{i}B_\mathrm{m}}{4\pi^2}\sum_{\nu=0}^{\nu_{\textrm{max}}}(2-\delta_{\nu 0})\bigg{[}\epsilon_{i}\sqrt{\epsilon_{i}^2-M_{i,\nu}^2}+M_{i,\nu}^2\textrm{ln}\bigg{(}\frac{\sqrt{\epsilon_{i}^2-M_{i,\nu}^2}+\epsilon_{i}}{M_{i,\nu}}\bigg{)}\bigg{]},
\end{eqnarray}
where $\epsilon_i$ is the Fermi energy for particle type $i$, $M_{i,\nu}=\sqrt{m_i^2+2\nu e_{i}B_{m}}$ is the quark effective mass in the presence of a magnetic field, $p_{i,\nu}$ is the maximum value of $p_z$ for the energy level $\nu$, satisfying the relation $\epsilon_{i}=\sqrt{p_{i,\nu}^{2}+M_{i,\nu}^2}$. From the positive value requirement on the Fermi momentum we can determine the upper bound $\nu_{\textrm{max}}$ of the summation index $\nu$

\begin{equation}\label{VMAX}
  \nu\leq\nu_{\textrm{max}}\equiv \textrm{int}\left[\frac{\epsilon_{i}^{2}-m_{i}^{2}}{2e_{i}B_\mathrm{m}}\right],
\end{equation}
where $\textrm{int}[A]$ means taking the integer part of $A$.

For a Fermi gas system, the number density is given by
\begin{equation}
 n_{i}=\frac{g_{i}}{(2\pi)^3}\int \mbox{d}^3 \vec{p}.
\end{equation}

In the presence of a magnetic field, according to the rule in Eq. (\ref{nu}), the expression of the number density can be apparently given as
\begin{eqnarray}\label{NIP}
 n_{i}&=&\frac{g_{i}e_{i}B_\mathrm{m}}{2\pi^2}\sum_{\nu=0}^{\nu_{\textrm{max}}}(2-\delta_{\nu 0})p_{i,\nu}\\\label{n2}
 &=&\frac{g_{i}e_{i}B_{m}}{2\pi^2}\sum_{\nu=0}^{\nu_{\textrm{max}}}(2-\delta_{\nu 0})\sqrt{\epsilon_{i}^{2}-M_{i,\nu}^2}\label{equE}.
\end{eqnarray}

In order to obtain the equation of states and check the thermodynamic consistency, we use the thermodynamic relation between pressure and the chemical potentials $\mu_i$:
\begin{equation}\label{P}
 P=-E+\sum_{i}\mu_{i}n_{i},
\end{equation}
which is valid for arbitrary infinitely large system. The chemical potentials are connected to the energy density by
\begin{equation}\label{relation}
 \mbox{d}E=\sum_{i}\mu_{i} dn_{i},
\end{equation}
where
\begin{equation}\label{E}
 \mbox{d}E=\sum_i \mbox{d}E_i=\sum_i \left(\frac{\partial E_i}{\partial \epsilon_i} \mbox{d}\epsilon_i+\frac{\partial E_i}{\partial m_i} \mbox{d} m_i \right),
\end{equation}

\begin{equation}\label{ni}
 \mbox{d} n_i=\frac{\partial n_i}{\partial \epsilon_i} \mbox{d} \epsilon_i+\frac{\partial n_i}{\partial m_i} \mbox{d} m_i,
\end{equation}

\begin{equation}\label{mi}
 \mbox{d} m_i=\sum_j \frac{\partial m_i}{\partial n_j} \mbox{d} n_j.
\end{equation}

Using these relations, we finally have
\begin{equation}\label{mui2}
 \mu_{i}=\epsilon_i+\mu_\mathrm I,
\end{equation}
with

\begin{equation}\label{ad}
\mu_\mathrm {I}=-\frac{m_\mathrm I}{9 n}\sum_{j}\frac{g_j e_j B_\mathrm{m}}{2 \pi^2 }\sum_{\nu=0}^{\nu_{\textrm{max}}}(2-\delta_{\nu 0})m_j\textrm{ln}\bigg{(}\frac{\epsilon_{j}+\sqrt{\epsilon_{j}^2-M_{j,\nu}^2}}{M_{j,\nu}}\bigg{)},
\end{equation}
where $m_\mathrm {I}=\frac{D}{n^{1/3}}$ is the interactive part in the mass scaling. We should notice that because all particle masses do not depend on the density of electrons, i.e. $\frac{\partial m_j}{\partial n_e}=0$, the summation is just over $u,d,s$ and the term as a whole is independent of quark flavors.
Substituting Eq. (\ref{EIY}), Eq. (\ref{n2}), Eq. (\ref{mui2}), Eq. (\ref{ad}) into Eq. (\ref{P}) gives
\begin{equation}\label{pressure}
P=-\Omega_0+\delta P,
\end{equation}
with
\begin{equation}
\Omega_0=-\sum_i \frac{g_i e_i B_\mathrm m}{4 \pi^2}\sum_{\nu=0}^{\nu_{\textrm{max}}}(2-\delta_{\nu 0})\bigg{\{}\epsilon_{i}(\epsilon_{i}^2-M_{i,\nu}^2)^{1/2}-M_{i,\nu}^2 \textrm{ln}\bigg{[}\frac{\epsilon_{i}+\sqrt{\epsilon_{j}^2-M_{j,\nu}^2}}{M_{i,\nu}}\bigg{]}\bigg{\}},
\end{equation}
and
\begin{equation}
\delta P=\sum_{i}n_i \mu_\mathrm I=3 \mu_\mathrm I n=-\frac{m_\mathrm I}{3}\sum_{j}\frac{g_j e_j B_m}{2 \pi^2 }\sum_{\nu=0}^{\nu_{\textrm{max}}}(2-\delta_{\nu 0})m_j\textrm{ln}\bigg{(}\frac{\epsilon_{j}+\sqrt{\epsilon_{j}^2-M_{j,\nu}^2}}{M_{j,\nu}}\bigg{)},
\end{equation}
where $\Omega_0$ is the thermodynamical potential density for a free-particle system in the presence of a magnetic field.

Comparing Eq. (\ref{pressure}) to constant-mass case where
\begin{equation}
P=-\Omega_0,
\end{equation}
we notice that the term $\delta P$, introduced by the density dependence of quark masses, guarantees the Hughenoltz-Van Hove theorem.

\section{Properties of mSQM \label{sec:posqm}}

In this section, we carry out a numerical study to investigate the relevant thermodynamical quantities for mSQM, taking into account $\beta$ equilibrium and charge neutrality. As usually done, we consider mSQM as a mixture of $u,d,s$ quarks and electrons. The weak equilibrium condition gives
\begin{equation}\label{AA}
\mu_u+\mu_e=\mu_d,
\end{equation}
\begin{equation}\label{BB}
\mu_d=\mu_s.
\end{equation}

Considering Eq. (\ref{mui2}), we have
\begin{equation}\label{equA}
\epsilon_{u}+\epsilon_{e}=\epsilon_{d},
\end{equation}
and
\begin{equation}\label{equB}
\epsilon_{d}=\epsilon_{s}.
\end{equation}

The charge neutrality condition gives
\begin{equation}\label{equC}
\frac{2}{3}n_u-\frac{1}{3}n_d-\frac{1}{3}n_s-n_e=0,
\end{equation}
and the baryon number density is given by
\begin{equation}\label{equD}
n=\frac{1}{3}(n_u+n_d+n_s).
\end{equation}

Eqs. (\ref{equA}), (\ref{equB}), (\ref{equC}) and (\ref{equD}), together with Eq. (\ref{equE}), form the full set of self-consistency equations for finding the Fermi energy for quarks and electrons. We solve these equations numerically to obtain the Fermi energies of each particle type at a given density $n$ and for different values of $B_m$, and then evaluate the thermodynamical quantities of the MSQM system.

\begin{figure}
  \includegraphics[width=10cm]{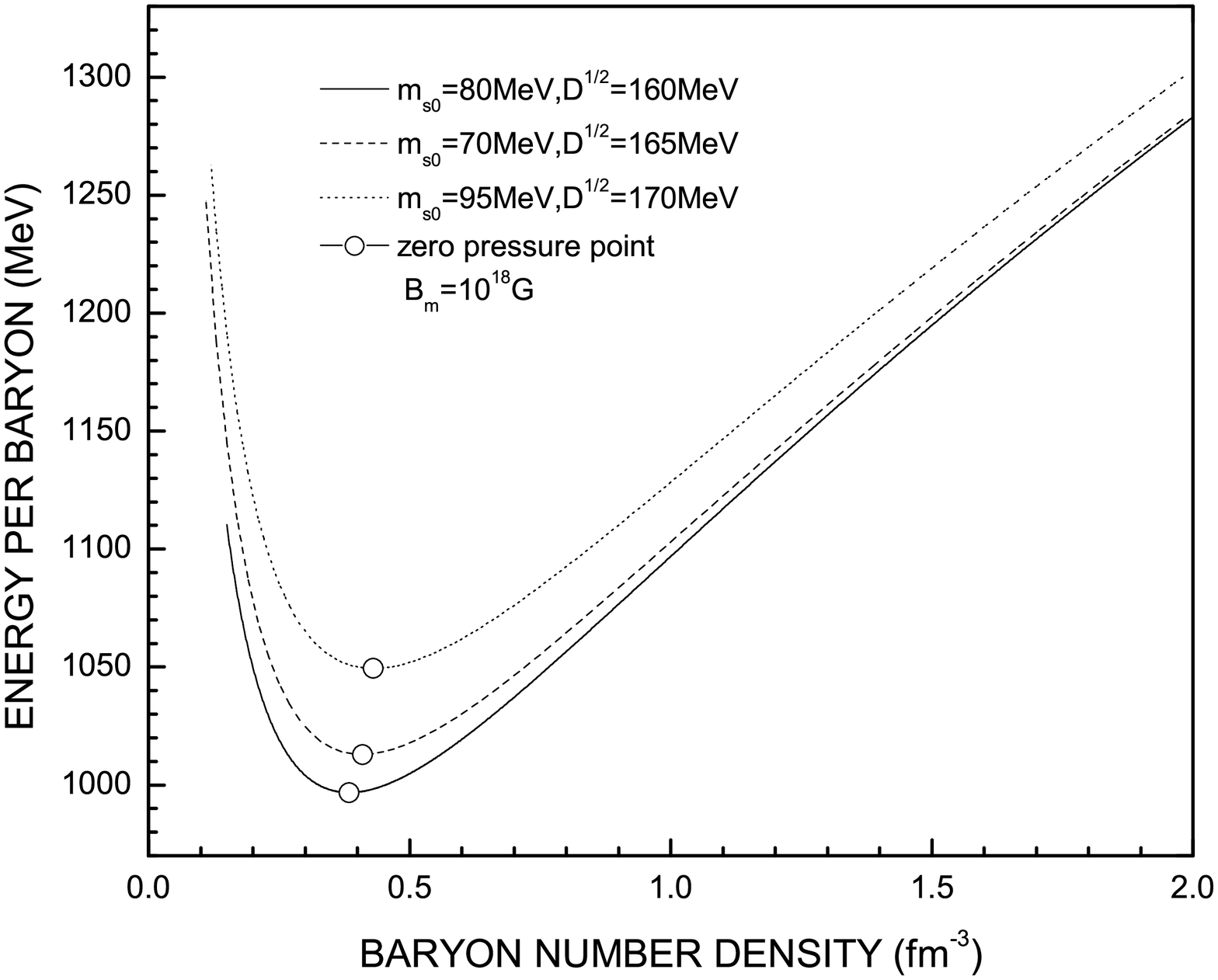}\\
  \caption{Energy per baryon of strange quark matter in the present model. The energy minimum is located exactly at the same point of the zero pressure.}\label{fig1}
\end{figure}

In Fig. \ref{fig1}, the energy per baryon of MSQM is shown as functions of different parameter sets. For each parameter set the pressure is zero at the minimum the energy. In fact from Eq. (\ref{P}) and Eq. (\ref{relation}) one can obtain $P=n^2\frac{d(E/n)}{dn}$. This relation guarantees the thermodynamic consistency. In the present paper, we assume the magnetic field to be directed along the $z$ axis and is a constant. If one allows the magnetic field to vary with the density, one should add to the chemical potential another new term. The discussion about the details of the magnetic field variation with the baryon number density is beyond this paper.

\begin{figure}
  \includegraphics[width=10cm]{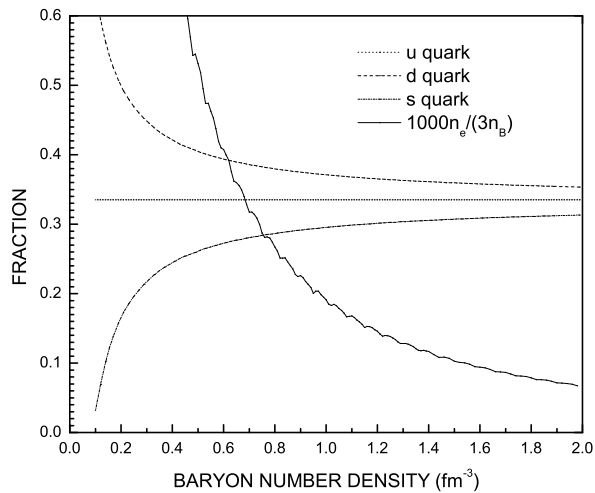}\\
  \caption{Quark fraction vs. baryon number density for $D^{1/2}=160$ MeV, $m_{s0}=80$ MeV and $B_m=10^{18}$ G.}\label{fig2}
\end{figure}

In Fig. \ref{fig2}, we plot the quarks fractions, i.e., $n_u/(3n), n_d/(3n), n_s/(3n)$, and the $10^3$ times the electron fraction, $1000n_e/(3n)$, versus the baryon number density for $D^{1/2}=160$ MeV and $m_{s0}=80$ MeV. The magnetic field strength is $B_\mathrm{m}=10^{18}$ G. The fraction of up quarks is always about one third. The fraction of down quarks decreases with increasing density while the fraction of strange quarks increases with increasing density. Both fractions approach one third when the baryon number density is large enough. The figure indicates that the fraction of electrons is very small and it decreases with increasing baryon number density. The ladder-like shape is introduced by the quantized laudau levels. We can expect that when the baryon number density increases to large enough value even the current mass of strange quark matter can be ignored and all three kinds of quarks can be treated equivalently. This is easy to understand. For fermions at zero temperature, the states below Fermi energy are all occupied. Adding particles into the system is just enlarge the Fermi energy, which is equivalent to magnify the Fermi momentum. At the same time the effective mass decreases with density. When the Fermi momentum is sufficiently large, the masses of most particles are ignorable compared to their momenta and we can treat them as if their effective masses are zero.

\begin{figure}
  \includegraphics[width=10cm]{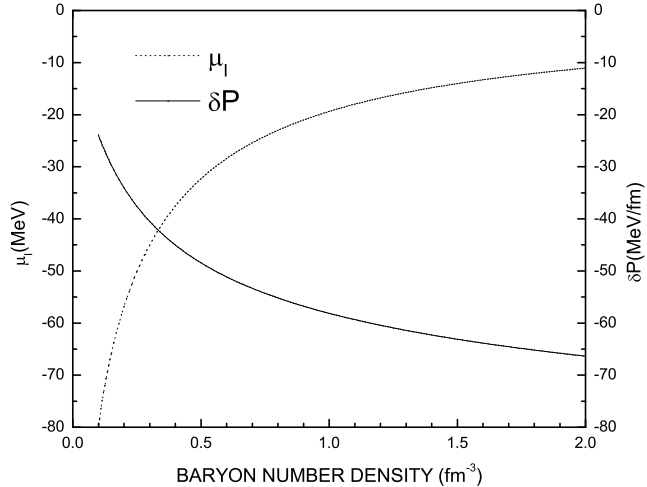}\\
  \caption{The terms introduced by the density dependence of quark masses for pressure and chemical potential versus baryon number density. $D^{1/2}=160$ MeV, $m_{s0}=80$ MeV and $B_\mathrm{m}=10^{18}$ G.}\label{fig3}
\end{figure}

To explicitly demonstrate the properties of $\mu_\mathrm {I}$ and $\delta P$, we plot them in Fig. \ref{fig3}. The additional term of the chemical potential tends to reduce the total value, but the effects gradually weaken when the baryon number density increases. For $\delta P$, it is a negative number and the curve shows a increasing tendency of the absolute value with baryon number density. From this set of curves we find that for respective type of particles the interactive part decreases with increasing the particle number density. When the density reach a sufficiently large value, the interactive effect can be ignored and the chemical potentials get close to the Fermi energy, which means that we can treat each type of fermions as free particles. But when dealing with the overall effect, the pressure, the interactive effect introduced by the dependence of masses on particle density, always exists and the effect of additional term reduces the pressure of the system.

\begin{figure}
  \includegraphics[width=10cm]{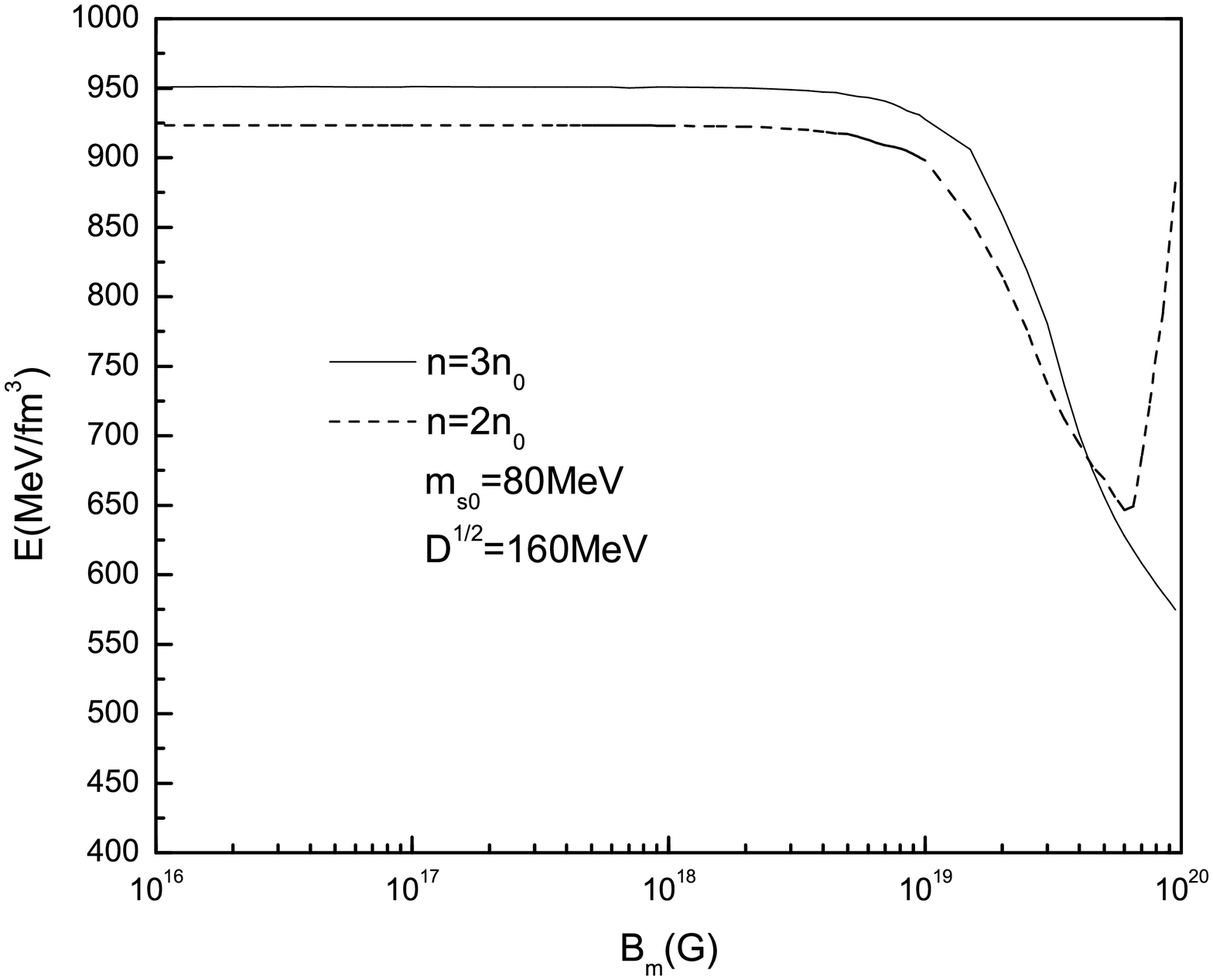}\\
  \caption{Energy density of mSQM as a function of the magnetic field strength. The parameter pair $(m_{s0},D^{1/2})$ in MeV is $(80,160)$. The solid line is for $n=3n_{0}$ and the dash line is for $n=2n_0$  ($n_{0}=0.16$ $\textrm{fm}^{-3}$).}\label{fig4}
\end{figure}

\begin{figure}
  \includegraphics[width=10cm]{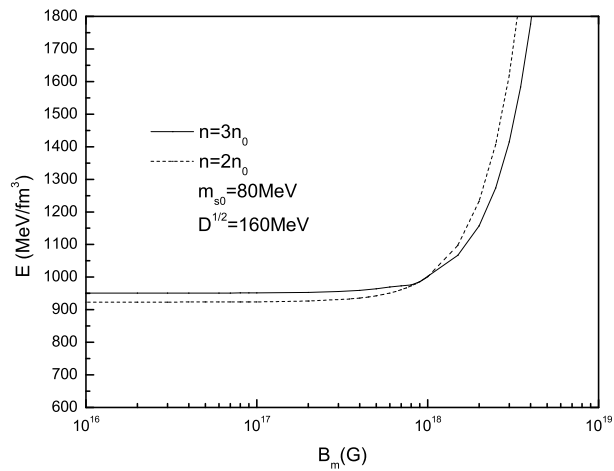}\\
  \caption{Energy density plus that of the magnetic field as a function of magnetic field strength. The parameter pair $(m_{s0},D^{1/2})$ in MeV is $(80,160)$. The solid line is for $n=3n_{0}$ and the dash line is for $n=2n_0$  ($n_{0}=0.16$ $\textrm{fm}^{-3}$).}\label{fig5}
\end{figure}

In Fig. \ref{fig4}, we show how the energy density of mSQM varies with the magnetic field strength at given densities $n=2n_0$ (dotted curve) and $n=3n_0$ (solid curve). For small $B_\mathrm{m}$, it is very obvious that the energy density of mSQM is nearly constant. When the magnetic field strength exceeds a critical value, which is about $3\times10^{18}$ G for $n=2n_0$, the energy density begins to decrease, until a minimum is reached. The minimum depends on the density. It is $6\times10^{19}$ G for $n=2n_0$. If one would hope to include the pure magnetic field term $B_\mathrm{m}^2/2$, as shown in Fig. \ref{fig5}, the minimum disappears and the total energy density increases monotonously. This is because the magnetic field itself becomes dominant at extremely strong strength.


\begin{figure}
  \includegraphics[width=10cm]{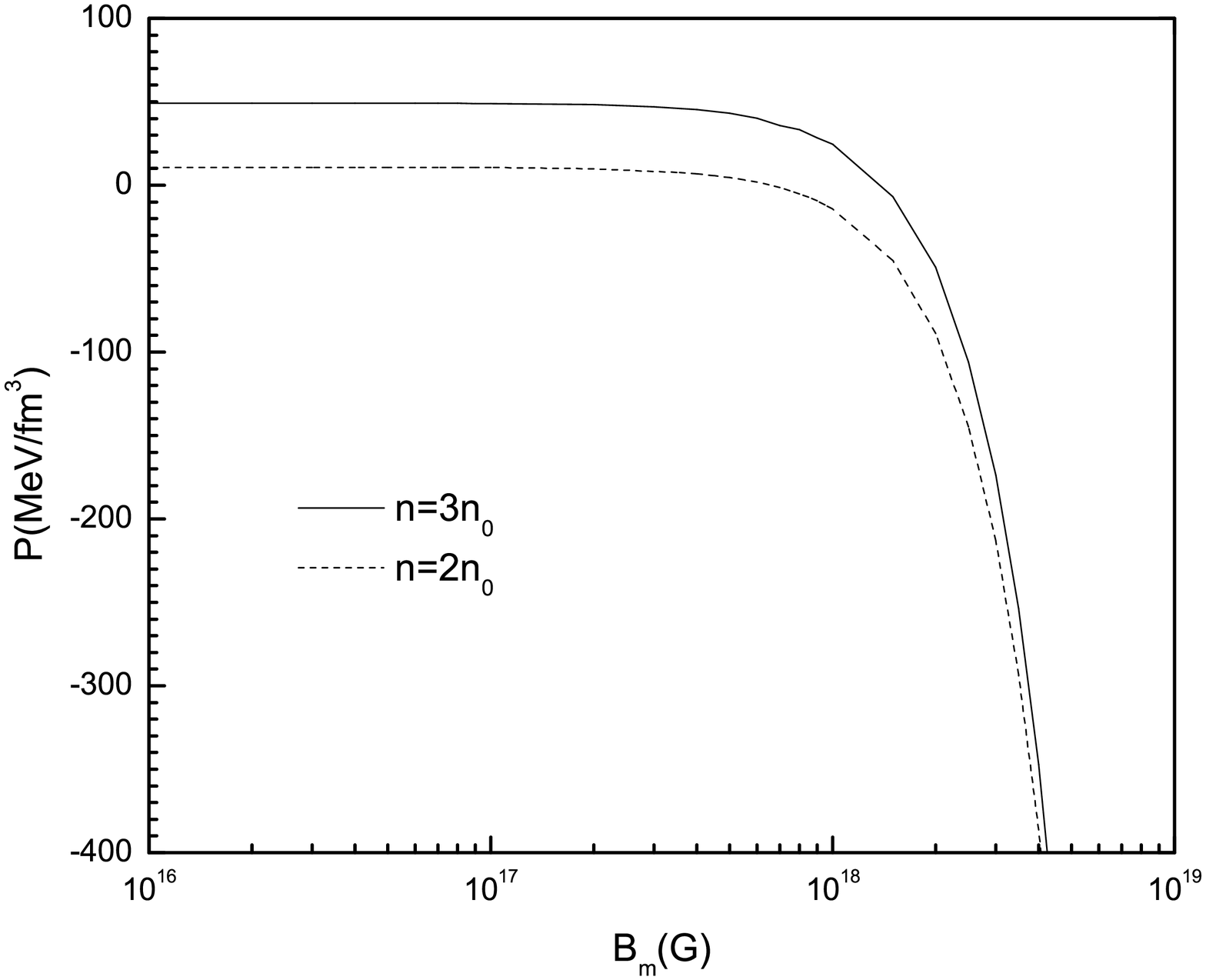}\\
  \caption{Pressure in mSQM at zero temperature as functions of the magnetic field strength for $m_{s0}=80$ MeV and $D^{1/2}=160$ MeV.}\label{fig6}
\end{figure}

In Fig. \ref{fig6} we show the pressure as a function of the strength of magnetic field for $n=3n_0$ (solid line) and $n=2n_0$ (dash line). At the beginning the pressure stays nearly constant and starts to decrease apperantly at $2\times10^{18}$ G. Since we using the CDDM model to investigate the properties of mSQM, we have to  assume the pressure to be isotropic because the Tolman-Oppenheimer-Volkoff (TOV) equations as usual method for finding mass-radius relationship only fits spherically symmetrical and static compact star. In Ref. \cite{Isayev2013}, the authors applied anisotropic pressure to strange quark matter system. The properties of the pressure in present article is in accordance with the longitudinal pressure in Ref. \cite{Isayev2013}.

\begin{figure}
  \includegraphics[width=10cm]{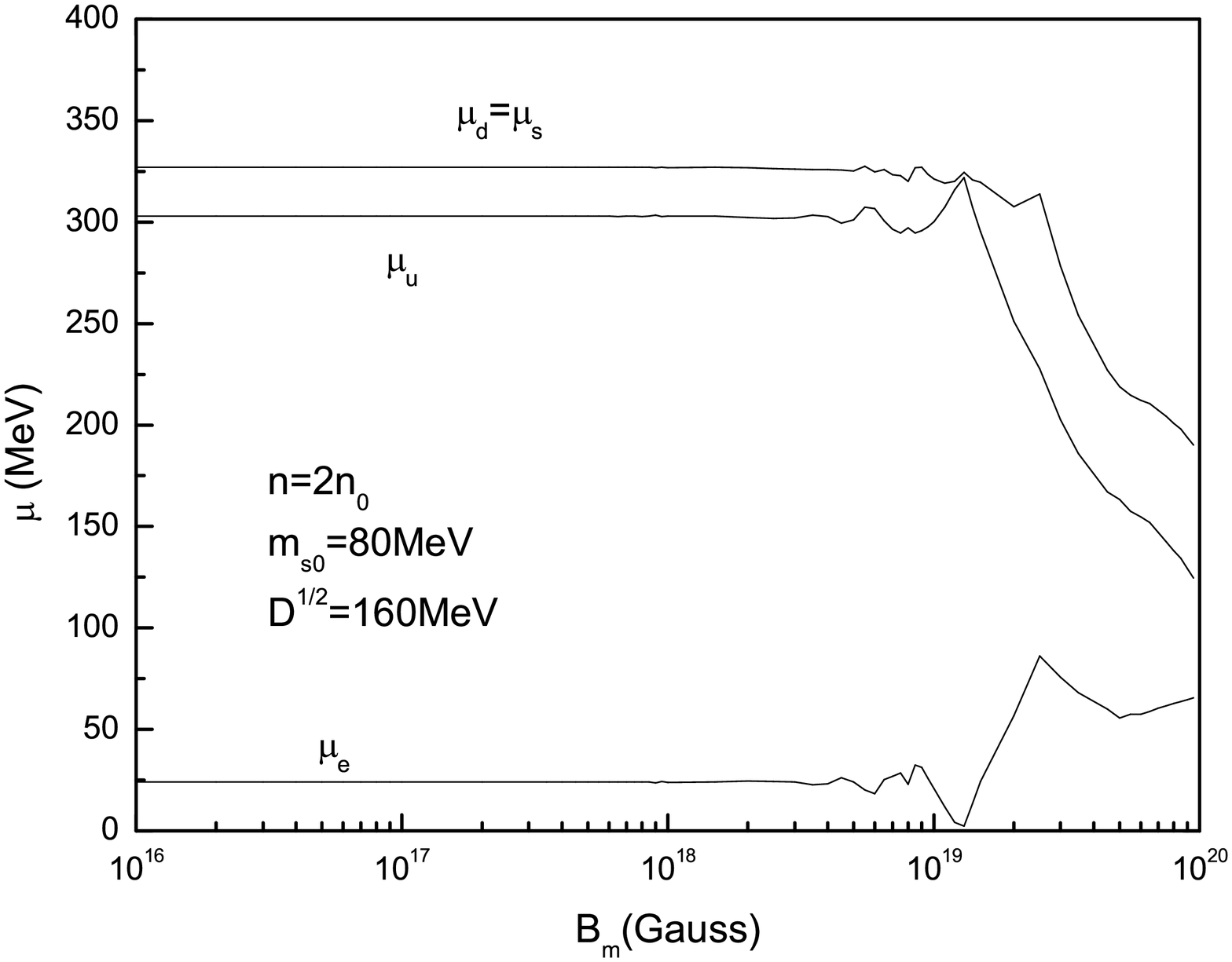}\\
  \caption{Chemical potentials as function of the magnetic field strength at zero temperature for $n=2n_0$.}\label{fig7}
\end{figure}

\begin{figure}
  \includegraphics[width=10cm]{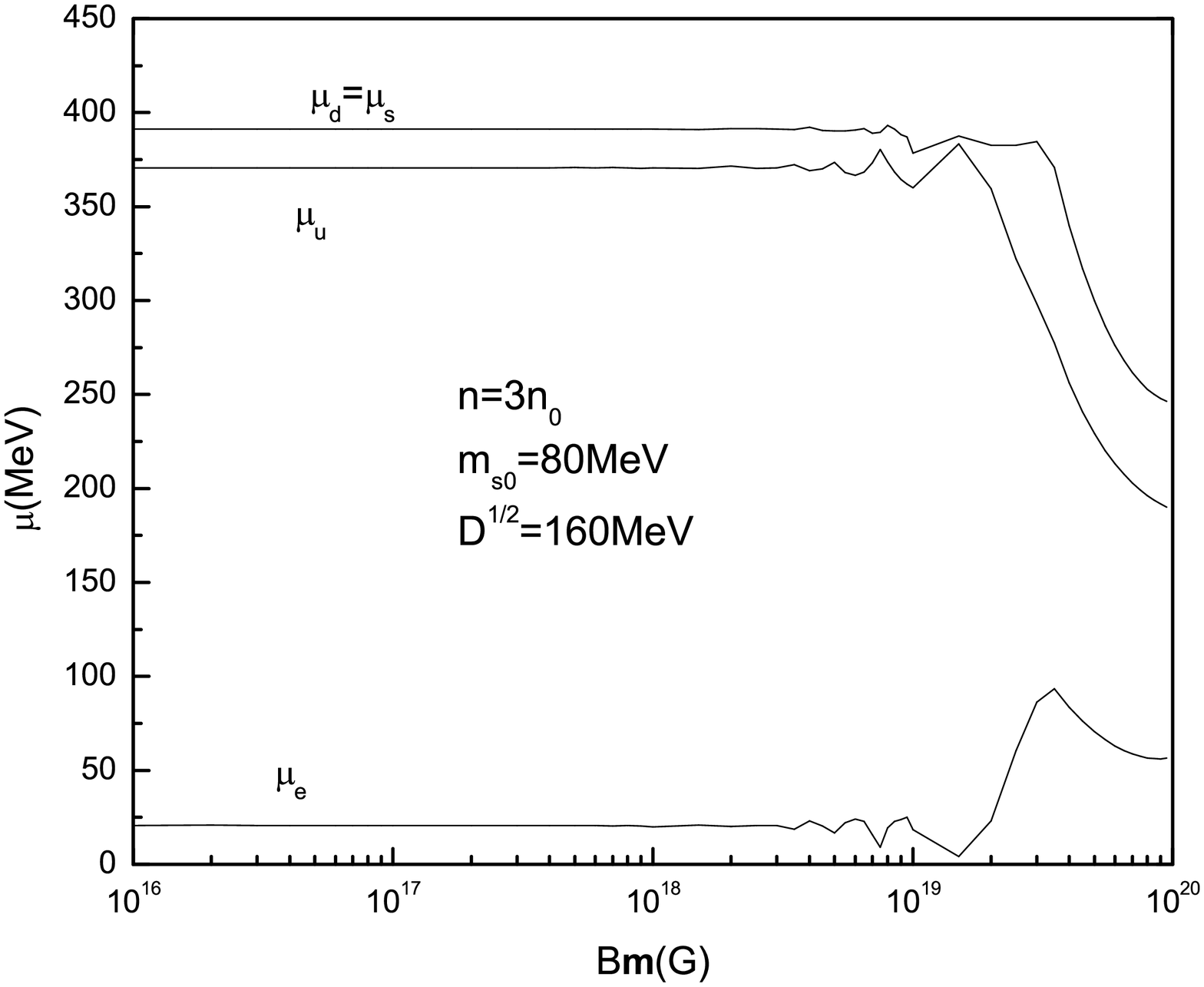}\\
  \caption{Chemical potentials as function of the magnetic field strength at zero temperature for $n=3n_0$.}\label{fig8}
\end{figure}

Fig. \ref{fig7} and Fig. \ref{fig8} show the chemical potentials as functions of the magnetic field for $n=2n_0$ and $n=3n_0$ respectively. The chemical potentials keep approximately constant with increasing magnetic field for $B_\mathrm{m}\leq10^{18}$ G. The apparent laudau oscillation of chemical potential appear between $3\times10^{18}$ G $\sim$ $2\times10^{19}$ G. At $B_\mathrm{m}\geq3\times10^{19}$ G quark chemical potentials decrease with magnetic field. We notice that in both figures the chemical potential of electron dramatically climbs from initial relatively small value to its summit (about $80$ MeV) in the interval $2\times10^{19}$ G$\leq B_\mathrm{m} \leq3\times10^{19}$ G, then decreases to a relatively high and even stage (about $60$ MeV). We have not encountered negative chemical potential for electron which contrasts with the situation in Ref. \cite{Isayev2013}.

\section{Mass-radius relation of magnetized strange quark stars \label{Mr}}
\begin{figure}
  \includegraphics[width=10cm]{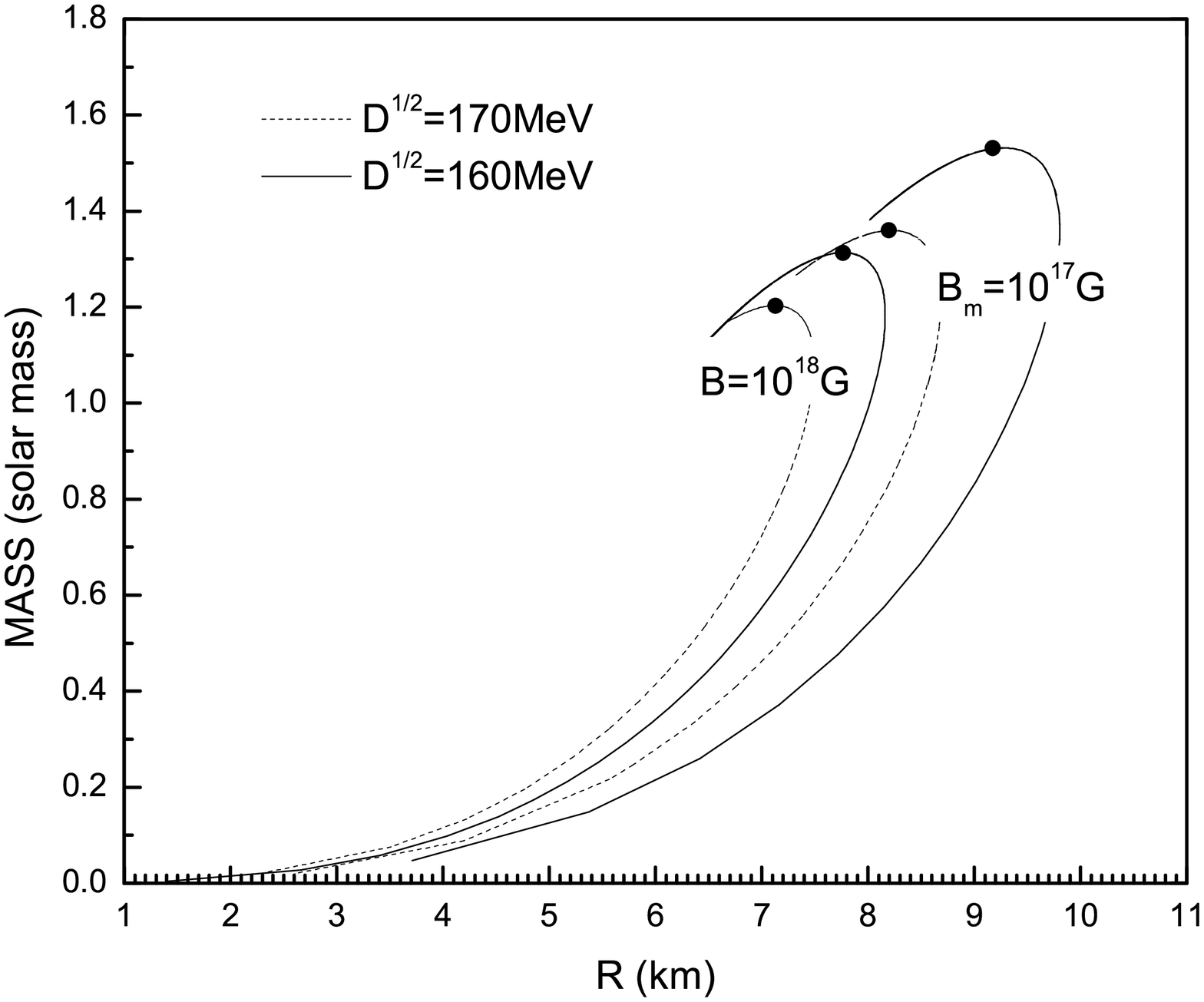}\\
  \caption{The mass-radius relation of SQS at different magnetic field $B_m$ with $D^{1/2}=160$ MeV (solid lines) and $D^{1/2}=170$ MeV (dash lines). The maximum masses on are marked by full dots.}\label{fig9}
\end{figure}

It has been speculated that the currently named neutron stars may in fact be strange quark stars \cite{Benvenuto1991}, a family of compact stars consisting completely of deconfined $u,d,s$ quarks. The structure of strange quark stars has attracted plenty of  researchers. We investigate the mass-radius relation of strange quark stars in ordinary phase in the framework of the new EOS we obtained in the preceding section.

We follow the method that numerically solving Toman-Oppenheimer-Volkoff (TOV) euqations when fixing a central pressure $P_c$, and obtain a mass-radius relation by continuously varying the central pressure. For a concrete description of the solving process, one may refer to Ref. \cite{Peng2000}. Since the method only fits spherically symmetrical and static compact star, we have assume the pressure to be isotropic in the entire article. From Sec. \ref{sec:posqm} we find that this assumption functions well at $B_\mathrm{m}\lesssim3\times10^{18}$ G. We alter parameter $D$ and magnetic field strength $B_\mathrm{m}$ to see how can the model and the magnetic field affect the structure of strange quark stars. The results have been showed in Fig. \ref{fig9}. The full dots represent the maximum mass for each line.

We can see that enhanced $B_\mathrm{m}$ or parameter $D$ could reduce the maximum mass. According to Fig. \ref{fig5} and Fig. \ref{fig6}, enhancing the magnetic field strength will increases energy density and decreases pressure which means softer EOS. Therefore the present EoS is unstable to produce the maximum star mass as large as two times the solar mass ($2 M_\odot$) \cite{Demorest2010,Antoniadis}. This is due to the fact that the quark mass scaling used in the present calculations includes only the confinement interaction effects whose contribution to the pressure is negative, while the perturbative interactions become important at high density. Therefore, it is meaningful to deduce a quark mass scaling considering both the confinement and perturbative effects \cite{Antoniadis}, which is an urgent forthcoming task in the near future.

\section{Summary \label{conclusion}}
We have extended the CDDM model with a cubic root mass scaling to study the properties of strange quark matter in the presence of a strong magnetic field. Our thermodynamic treatment automatically guarantees the self-consistency. It is found that at high density quarks of different kinds can be treated equivalently for dynamic properties and individual particle acts like free particle while the overall effect introduced by the density dependence of quark masses always exist. The magnetic field will reduce the energy density of pure magnetized strange quark matter at certain range of the field strength. At $B_\mathrm{m}\leq10^{18}$ G, magnetic field affects the properties of the system slightly. At $10^{18}$ G$\leq B_\mathrm{m} \leq 10^{19}$ G, Laudau oscillation appears in chemical potential and the pressure dramatically decreases. At $B_\mathrm{m}\approx4\times10^{18}$ G the energy density of pure magnetic field becomes comparable with, and finally much large than that of the pure magnetized SQM. With the obtained EoS, we study the mass-radius relation of quark stars. It is found that one can not produce a pure quark star with mass as large as $2 M_\odot$ considering only the confinement interaction in the quark mass scaling. A quark mass scaling with both the confinement and perturbative interactions are needed.


\begin{acknowledgments}
The authors would like to thank support from the National Natural Science Foundation of China (Grant No. 11135011) and by the CAS key project Y12A0A0012.
\end{acknowledgments}


\begin{thebibliography}{99}


\bibitem{Bodmer1971}
  A. R. Bodmer,
Phys.\ Rev.\ D {\bf 4}, 1601 (1971).





\bibitem{Terazawa1979}
  H. Terazawa,
INS-Report 336, Univ. of Tokyo (1979).

\bibitem{Witten1984}
  E. Witten,
Phys.\ Rev.\ D {\bf 30}, 272 (1984).


\bibitem{Farhi1984}
  E. Farhi and R. L. Jaffe,
Phys.\ Rev.\ D {\bf 30}, 2379 (1984).


\bibitem{Jaffe1974}
  A. Chodos, R. L. Jaffe, K. Johnson, C. B. Thorn, and V. F. Weiskopf,
Phys.\ Rev.\ D {bf 9}, 3471 (1974).

\bibitem{Hatsuda1987}
  T. Hatsuda,
  Mod.\ Phys.\ Lett.\ A {\bf 2}, 805 (1987).

\bibitem{Sato1987}
 K. Sato and H. Suzuki,
 Phys.\ Rev.\ Lett.\  {\bf 58}, 2722 (1987).

\bibitem{Chakrabarty1996}
  S. Chakrabarty,
  Phys.\ Rev.\ D {\bf 54}, 1306 (1996).

\bibitem{Duncan1992}
  R. Duncan and C. Thompson,
  Astron.\ J. {\bf 32}, L9 (1992).

\bibitem{Kouveliotou1988}
 C. Kouveliotou {\it et al},
 Nature {\bf 393}, 235 (1988).

\bibitem{Tatsumi2006}
  T. Tatsumi, T. Maruyama, E. Nakano, and K. Nawa,
  Nucl.\ Phys.\ A {\bf 774}, 827 (2006).


\bibitem{Rojas2005}
  A. P\'{e}rez Mart\'{\i}nez, H. P\'{e}rez Rojas, H. J. Mosquera Cuesta, M. Boligan, and M. G. Orsaria,
  Int.\ J.\ Mod.\ Phs.\ D {\bf 14}, 1959 (2005).

\bibitem{Rojas2007}
  A. P\'{e}rez Mart\'{\i}nez, H. P\'{e}rez Rojas, H. J. Mosquera Cuesta, and M. G. Orsaria,
  Int.\ J.\ Mod.\  Phys.\ D {\bf 16}, 255 (2007).


\bibitem{Chaichian2000}
  M. Chaichian, S. S. Masood, C. Montonen, A. P\'{e}rez Mart\'{\i}ez, and H. P\'{e}rez Rojas,
  Phys.\ Rev.\ Lett. {\bf 84}, 5261 (2000).


\bibitem{Rojas2003}
  A. P\'{e}rez Mart\'{\i}ez, H. P\'{e}rez Rojas, and H. J. Mosquera Cuesta,
  Eur.\ Phys.\ J.\ C {\bf 29}, 111 (2003).


\bibitem{Mosquera2005}
  R. Gonz\'{a}lez Felipe, H. J. Mosquera Cuesta, A. P\'{e}rez Mart\'{\i}ez, and H. P\'{e}rez Rojas,
  Chin.\ J.\ Astron.\ Astronphys. {\bf 5}, 399 (2005).

\bibitem{Ebert2003}
 D. Ebert and K. G. Klimenko,
 Nucl.\ Phys.\ A {\bf 728}, 203 (2003).


\bibitem{Frolov2010}
  I. E. Frolov, V. C. Zhukovsky, and K. G. Klimenko,
  Phys.\ Rev.\ D {\bf 82}, 076002 (2010).

\bibitem{Fayazbakhsh2011}
  S. Fayazbakhsh and N. Sadooghi,
  Phys.\ Rev.\ D {\bf 83}, 025026 (2011).


\bibitem{Menezes2009}
D. P. Menezes, M. Benghi Pinto, S. S. Avancini, and C. Provid\^{e}ncia,
Phys.\ Rev.\ C {\bf 80}, 065805 (2009).


\bibitem{Avancini2011}
  S. S. Avancini, D. P. Menezes, and C. Provid\^{e}ncia,
  Phys.\ Rev.\ C {\bf 83}, 065805 (2011).

\bibitem{Mizher2010}
  A. J. Mizher, M. N.  Chernodub, and E. S. Fraga,
  Phys.\ Rev.\ D {\bf 82}, 105016 (2010).



\bibitem{Rabhi2011}
A. Rabhi and C. Provid\^{e}ncia,
Phys.\ Rev.\ C {\bf 83}, 055801 (2011).



\bibitem{Walecka1995}
  J. D. Walecka,
  Oxford Stud.\ Nucl.\ Phys. {\bf 16}, 1 (1995).


\bibitem{Henley1990}
  E. M Henley, and H. M\"{u}ther,
  Nucl.\ Phys.\ A {\bf 513}, 667 (1990).


\bibitem{Brown1991}
  G. E. Brown and M. Rho,
  Phys.\ Rev.\ Lett. {\bf 66}, 2720 (1991),


\bibitem{Cohen1992}
  T. D. Cohen, R. J. Furnstahl, and D. K. Griegel,
  Phys.\ Rev.\ Lett. {\bf 67}, 961 (1991); Phys.\ Rev.\ C {\bf 45}, 1881 (1992).


\bibitem{Schertler1997}
  K. Schertler, C. Greiner, and M. H. Thoma,
  Nucl.\ Phys.\ A {\bf 616}, 659 (1997).

\bibitem{Buballa1999}
M. Buballa and M. Oertel,
Phys.\ Lett.\ B {\bf 457}, 261 (1999).

\bibitem{Wen2012}
X. J. Wen, S. Z. Su, D. H. Yang, and G. X. Peng,
Phys.\ Rev.\ D {\bf 86}, 034006 (2012).

\bibitem{Fowler1981}
G. N. Fowler, S. Raha, and R. M. Weiner,
Z.\ Phys.\ C {\bf 9}, 271 (1981).

\bibitem{Chakrabarty1989}
S. Chakrabarty, S. Raha and B. Sinha,
Phys.\ Lett.\ B {\bf 229}, 113 (1989).

\bibitem{Chakrabarty1991}
S. Chakrabarty,
Phys.\ Rev.\ D {\bf 43}, 627 (1991).

\bibitem{Chakrabarty1993}
S. Chakrabarty,
Phys.\ Rev.\ D {\bf 48}, 627 (1993).

\bibitem{Peng1999}
G. X. Peng, H. C. Chiang, J. J. Yang, L. Li, B. Liu,
Phys.\ Rev.\ C {\bf 61}, 015201 (1999).

\bibitem{Peng2000}
G. X. Peng, H. C. Chiang, B. S. Zhou, P. Z. Ning, and S. J. Luo,
Phys.\ Rev.\ C {\bf 62}, 025801 (2000).

\bibitem{Wen2005}
X. J. Wen, X. H. Zhong, G. X. Peng, P. N. Shen, N. Z. Ning,
Phys.\ Rev.\ C {\bf 72}, 015204 (2005).

\bibitem{Wang2000}
P. Wang,
Phys.\ Rev.\ C {\bf 62}, 015204 (2000).

\bibitem{Peng1997}
G. X. Peng, P. Z. Ning, and H. Q. Chiang,
Phys.\ Rev.\ C {\bf 56}, 491 (1997).

\bibitem{Zheng2004}
X. P. Zheng, X. W. Liu, M. Kang, and S. H. Yang,
Phys.\ Rev.\ C {\bf 70}, 015803 (2004).

\bibitem{Lugones2003}
G. Lugones and J. E. Horvath,
Int.\ J.\ Mod.\ Phys.\ D {\bf 12}, 495 (2003).

\bibitem{Wen2007}
X. J. Wen, G. X. Peng, and Y. D. Chen,
J.\ Phys.\ G: Nucl.\ Part.\ Phys. {\bf 34}, 1697 (2007).

\bibitem{Shen2007}
X. J. Wen, G. X. Peng, and P. N. Shen,
Int.\ J.\ mod.\ Phys.\ A {\bf 22}, 1649 (2007); G. X. Peng, X. J. Wen, and Y. D. Chen,
Phys.\ Lett.\ B {\bf 633}, 313 (2006).

\bibitem{Yao2006}
W. M. Yao \textit{et al.},
J.\ Phys.\ G: Nucl.\ Part.\ Phys. {\bf 33}, 1 (2006).

\bibitem{Peng2008}
G. X. Peng, A. Li, and U. Lombardo,
Phys.\ Rev.\ C {\bf 77}, 065807 (2008).

\bibitem{Laudau1965}
L. D. Laudau and E. M Lifshitz,
\textit{Quantum Mechanics} (Pergamon Press, New York, 1965).

\bibitem{Isayev2013}
A. A. Isayev and J. Yang,
J.\ Phys.\ G: Nucl.\ Part.\ Phys. {\bf 40}, 035105 (2013).

\bibitem{Benvenuto1991}
O. G. Benvenuto, J. E. Horvath, and H. Vucetich,
Int.\ J.\ Mod.\ Phys.\ A {\bf 6}, 4769 (1991);
C. Alcock and A. Olinto,
Annu.\ Rev.\ Nucl.\ Part.\ Sci. {\bf 38}, 161 (1988).

\bibitem{Demorest2010}
P. Demorest, T. Pennucci, S. Ransom, M. Roberts, and J. Hessels,
Nature (London) {\bf 467}, 1081 (2010).

\bibitem{Antoniadis}
J. Antoniadis \textit{et al.},
Science {\bf 340}, 1233232 (2013).

\bibitem{Pengnucl2005}
Peng,
Nucl. Phys. A {\bf 747}, 25 (2005).


\end{thebibliography}
\end{document}